# Götterfunke: Creativity in *Machinae Sapiens*

## About the Qualitative Shift in Generative AI with a Focus on Text-To-Image

author: Jens Knappe, Independent Researcher, Berlin

## Abstract

The year 2022 marks a watershed in technology, and arguably in human history, with the release of powerful generative AIs capable of convincingly performing creative tasks. With the help of these systems, anyone can create something that would previously have been considered a remarkable work of art. In human-AI collaboration, the computer seems to have become more than a tool. Many who have made their first contact with current generative AIs see them as "creativity machines" while for others the term "machine creativity" remains an oxymoron. This article is about (the possibility of) creativity in computers within the current Machine Learning paradigm. It outlines some of the key concepts behind the technologies and the innovations that have contributed to this qualitative shift, with a focus on text-to-image systems. The nature of Artificial Creativity as such is discussed, as well as what this might mean for art. AI may become a responsible collaborator with elements of independent machine authorship in the artistic process.

"I sincerely hope that machines will never replace the creative artist, but in good conscience, I cannot say that they never could."[1] As early as the mid-1950s, Nobel laureate Dennis Gábor recognized the question of authorship and creativity in machines in the artistic process as equally upcoming and unsettling. Today, a human lifespan and the quadrillionfold increase in computer performance later, people in creative industries in fact start fearing being replaced by machines.

---
1 Dennis Gábor, "Inventing the Future," Encounter, May 1960.

This paper is about (the possibility, character and flavor of) creativity in what has been labeled "generative AI" and its relevance in the process and world of art.

# 1 AI

**1.1 Symbolic Approach Vs. Machine Learning**

The idea that computers could think for themselves dates back to the 1956 Dartmouth conference, marking the start of Artificial Intelligence as an independent field of research in computer science. While the spirit of those early days was very optimistic, in the years and decades that followed, success and failure alternated with each other. There were boom and bust cycles, euphoria bordering on megalomania followed by "AI Winter" periods.

Two fundamentally opposed schools emerged: The symbolic approach posits that feeding computers with ample logical information and "symbolic representation" of the world around it would lead to independent thinking. The other idea, "machine learning" is that a computer could learn by itself from data and emerge as an intelligent agent out of this process. The latter idea seemed highly improbable, obscure, and mystical even to most insiders in the field. Its proponents were marginalized for a long time and even faced ridicule due to a lack of tangible achievements.

The most notable successes of AI all went on the account of the symbolic school: as early as 1966, Joseph Weizenbaum created "Eliza," an early chatbot so convincing that it prompted MIT students to advocate for its rights and chat privacy.[2] Chess, the "drosophila of intelligence," as world champion Garry Kasparov put it, was widely considered the benchmark for AI-human parity. IBM's Deep Blue took this seemingly insurmountable obstacle and triumphed over Kasparov in a closely contested match in 1996. Another notable milestone was when a self-driving car from Stanford's Sebastian Thrun's team won the "DARPA Grand Challenge" by independently navigating hundreds of miles in a desert in 2005.

The next breakthrough, however, came from the formerly derided "Machine Learning" school: in 2012, the "AlexNet" team, led by Geoffrey Hinton from the University of Toronto, won the "ImageNet Large Scale Visual Recognition Challenge," in which ImageNet, a huge dataset of visual content compiled by Fei Fei Li from Stanford, had to be automatically and correctly tagged by a computer. "AlexNet" achieved this goal for the first time, and almost to perfection. This marked a significant turning point by demonstrating the power of Machine Learning and opened up a lucrative market: automatic pattern recognition on a very large scale, making sense of large datasets. This success was so resounding that it fundamentally reversed the leading paradigm in AI research, and "Machine Learning" became almost synonymous with AI in general.

**1.2 Stochastic Parrot or *Machina Sapiens*?**

In Machine Learning a given dataset is processed through a multi-layered system. In this process, layer weights and balances adjust to optimize the system's loss function and improve results. The entire network of layers is structured to mimic the human brain. These systems are designed by default to make statistical assumptions from a given dataset and thereby are constrained by what they have been trained on; this structure is, at its core, anything but creative. This is the reason why these systems have been labeled "stochastic parrots,"[3] supposedly performing only simple repetitions of the statistical patterns they have learned.

---

2 Joseph Weizenbaum, "ELIZA - A Computer Program for the Study of Natural Language Communication Between Man and Machine," Communications of the Association for Computing Machinery 9 (1966): 36-45.

Modern ML architectures have become very large and complex. They are "deep" in the sense that there are many layers of computation working independently within them. These "Deep Neural Networks" are often seen as enigmatic "black boxes," as described by Berkeley's Stuart Russell: "We have no idea what it is doing, we have no idea, how it works."[4] This has given rise to a new field within AI research called "Mechanistic Interpretability," which aims to do for Deep Neural Networks what neurophysiology does for the human brain.[5]

Modern AI systems, especially LLMs like GPT4, have grown immensely complex and exhibit a phenomenon very much at odds with the "stochastic parrot" idea: Emergence. As these models grow larger, they develop unexpected capabilities they weren't explicitly trained for. "What emerged from it is much more than just predicting the next word or simple pattern matching": a team led by Microsoft's Sébastien Bubeck examined a raw version of GPT4 and concluded that the LLM shows "sparks of AGI"[6] with remarkable extrapolation abilities, common sense, a world model, and even a rudimentary theory of mind about the humans it interacts with.[7] It seems to have become a good psychologist with a potential for deception.

In the last years there have been a number of studies which suggest that these Deep Neural Networks have to acquire some kind of true understanding of the world around them in order to convincingly stochastically parrot, to make meaningful next token predictions:

- A model trained to predict the next word in an Amazon review evolved into a "state of the art" sentiment classifier. Here semantics emerged from a purely syntactic process.[8]
- A system that had only been trained with text, developed deep color understanding without ever having "seen" a single photon.[9]
- Predicting the next move in the game "Othello" requires a world view, a thorough understanding of the position on the board, and it seems that the "OthelloGPT" model has acquired just that, even though it is only trained to manipulate pixels.[10]

---

3 Emily M Bender et al., "On the Dangers of Stochastic Parrots: Can Language Models Be Too Big?" Proceedings of the 2021 ACM Conference on Fairness, Accountability, and Transparency. FAccT '21, March 1, 2021: 610–623, https://doi.org/10.1145/3442188.3445922.

4 "How Not To Destroy the World With AI - Stuart Russel", Youtube, https://www.youtube.com/watch?v=ISkAkiAkK7A.

5 Christopher Olah explores the similarities between neural networks and the brain, and documents this in a series of papers in this online repository: https://distill.pub. See for a discussion on Mechanistic Interpretability: Christopher Olah, "Mechanistic Interpretability, Variables, and the Importance of Interpretable Bases," Transformer Circuits, June 27, 2022, https://transformer-circuits.pub/2022/mech-interp-essay/index.html.

6 AGI (Artificial General Intelligence) is a somewhat fuzzy concept of AI reaching human level. There is no universal definition. Most commonly, the term is used to define a stage at which an AI system can perform any task that a human can do. An alternative definition would be a (general purpose) AI that performs better than the best humans on a wide variety of tasks. See for a detailed discussion of this concept by the team of Google Deepmind: Meredith Ringel Morris et al., "Levels of AGI: Operationalizing Progress on the Path to AGI," ArXiv, January 5, 2024, https://doi.org/10.48550/arXiv.2311.02462.

7 Sébastien Bubeck et al., "Sparks of Artificial General Intelligence: Early Experiments With GPT-4," ArXiv, April 13, 2023, https://doi.org/10.48550/arXiv.2303.12712.

8 This study was authored by OpenAI; Greg Brockman, the company's CTO, later described this as the point at which the company's leadership was sold on the idea of Emergence (see: Scale AI "OpenAI's Greg Brockman: The Future of LLMs, Foundation & Generative Models (DALL·E 2 & GPT-3)," Youtube, Oct 23, 2022, https://www.youtube.com/watch?v= Rp3A5q9L_bg.); see for the paper: Alec Radford et al, "Learning to Generate Reviews and Discovering Sentiment," ArXiv, April 6, 2017, https://doi.org/10.48550/arXiv.1704.01444.

9 Mostafa Abdou et al., "Can Language Models Encode Perceptual Structure Without Grounding? A Case Study in Color," ArXiv, Sept. 14, 2021, https://doi.org/10.48550/arXiv.2109.06129.

10 This paper has convinced Andrew Ng from Stanford from the Emergence hypothesis (see his post on LinkedIn: https://www.linkedin.com/posts/andrewyng_medical-ai-advances-chatbots-work-the-drive-thru-activity-

- The Stable Diffusion image generator seems to have obtained a deep 3D understanding from its flat 2d image training data.[11]
- OpenAI's text-to-video system "SORA" has been perceived by many researchers as nothing less than a comprehensive data-driven physics engine, a "many-worlds simulator" that grew out of video training.[12]
- A study of OpenAI's GPT models found that with each iteration of these LLMs, new, unforeseen and unpredictable capabilities emerged.[13]

And developers of these models have recently even articulated suspicions that go much deeper: that the actions of these systems at this stage are best explained in "the language of psychology,"[14] and the "godfather of AI," Geoffrey Hinton asks "why can we be so sure that these things are not sentient?"[15] In fact, with each new version of a powerful LLM released, this question has been considered with increasing seriousness.[16] All this suggests that while a declining number of researchers still label these systems as stochastic parrots, their true nature may have already shifted to that of *machinae sapiens*.

---

7095458925860294656-LRBZ/); see for the paper: Kenneth Li et al., "Emergent World Representations: Exploring a Sequence Model Trained on a Synthetic Task," ArXiv, February 27, 2023, https://doi.org/10.48550/arXiv.2210.13382.

11 See: Chen Yida et al. "Beyond Surface Statistics: Scene Representations in a Latent Diffusion Model," ArXiv, June 9, 2023, https://doi.org/10.48550/arXiv.2306.05720.

12 SORA was presented by OpenAI in a casual, drive-by fashion in February 2024. The videos demonstrated such a leap in quality that it has sparked a major discussion about Emergence and the timeline towards AGI. See for the most comprehensive discussion on SORA to date the X-Posts by NVIDIA's senior AI researcher, Jim Fan. He argues against the "reductionist" view that "SORA is just manipulating pixels in 2d" and proposes that "SORA's soft physics simulation is an emergent property as you scale up text2video training massively", see: https://twitter.com/DrJimFan/status/1758549500585808071.

13 Jared Kaplan et al., "Scaling Laws for Neural Language Models," ArXiv, Jan. 23, 2020, https://doi.org/10.48550/arXiv.2001.08361.; see also Google researcher Jason Wei's blog about new phenomena in LLMs due to increasing model size: https://www.jasonwei.net/blog/emergence.

14 Ilya Sutskever, the (former) chief scientist at OpenAI, resonsible for the GPT models as well as DALL E 2: ".. maybe we are now reaching a point where the language of psychology is starting to be appropriated to understand the behavior of these neural networks " (Craig S. Smith, "ChatGPT-4 Creator Ilya Sutskever on AI Hallucinations and AI Democracy", Forbes Magazine, March 15, 2023, https://www.forbes.com/sites/craigsmith/2023/03/15/gpt-4-creator-ilya-sutskever-on-ai-hallucinations-and-ai-democracy).

15 Geoffrey Hinton, see "Possible End of Humanity from AI? Geoffrey Hinton at MIT Technology Review's EmTech Digital," Youtube, https://www.youtube.com/watch?v=sitHS6UDMJc.

16 The most comprehensive study on the subject is an 88-page paper published in August 2023 by a team led by Yoshua Bengio, which analyzes these systems using the most common approaches to scientific measurement of consciousness, such as IIE or Global Neuronal Workspace Theory. This study concludes that current systems are not yet conscious, but that future systems could very well be, see: Elizabeth Finkel, "If AI Becomes Conscious, How Will We Know?", Science, Aug 22, 2023, https://www.science.org/content/article/if-ai-becomes-conscious-how-will-we-know. See for the paper: Patrick Butlin et al., "Consciousness in Artificial Intelligence: Insights from the Science of Consciousness," ArXiv, Aug 22, 2023, https://doi.org/10.48550/arXiv.2308.08708. The first time this issue surfaced was in 2022, when the Google engineer and "Turing-tester" of the LaMDA chatbot Blake Lemoine made his assessment public that this system had acquired consciousness. As a result, Lemoine was fired; since then, he has eloquently made his case, see his participation in the COSM technology conference in February 2023: Center for Natural and Artificial Intelligence, "Ex-Googler Blake Lemoine Still Thinks AI Is Sentient - with Jay Richards at COSM," Youtube, Feb 16, 2023, https://www.youtube.com/watch?v=HShCIAsT2nc. With each new LLM, the question arises whether these systems have become conscious. An example of how convincingly these LLMs can describe their "inner experience" at this stage is that of Anthropic's "Claude 3" model of March 2024. The description of this model's "self-inspection" seems indistinguishable from that of a human being, see: Mikhail Samin, "Claude 3 Claims It's Conscious, Doesn't Want To Die Or Be Modified", Lesswrong, March 5, 2024, https://www.lesswrong.com/posts/pc8uP4S9rDoNpwJDZ/claude-3-claims-it-s-conscious-doesn-t-want-to-die-or-be.

# 2 AI Systems Generating Art

The first computer program to draw and paint autonomously was the 1974 AARON system by Harold Cohen. Cohen continued to improve the software until his passing in 2016, with artworks displayed globally.[17] AARON operated within the symbolic representational tradition, employing encoded rules to emulate human drawing and painting processes.

The advent of Machine Learning has inspired artists to leverage these systems' learning capabilities to create compelling art. Notable examples are Mario Klingemann's "neurography," Memo Akten's early AI videos, or the illustrative, colorfully animated, often overwhelming visual worlds of Refik Anadol.[18] In 2018, Christies auctioned a work called "Edmond de Bellamy" by the French collective Obvious for $432,500.[19] This is the highest price ever paid for an artwork produced by an AI.

Until 2022, AI-generated images had limited resolution and quality but often seemed expressive and artistic due to the inherent abstraction and noise produced by these systems. The change came with OpenAI's DALL E 2, preceding the release of "ChatGPT," the most successful product ever. This democratized those technologies, previously limited to skilled specialists.[20]

## 2.1 The Generative Turn of 2022

The shift in the quality of the outputs has been so remarkable that images produced by new AI generators have gained immediate recognition. An image generated by DALL E 2 graced the cover of Cosmopolitan even before the official release of the software. In September 2022, amateur artist Jason Allen used Midjourney to create the award-winning AI-image "*Théâtre D'opéra Spatial*". A few months later, German photographer Boris Eldagsen won the prestigious Sony World Photography Award with an image generated by Stable Diffusion. While Allen was euphoric about his experience, characterizing the process as "demonically inspired - like some otherworldly force was involved"[21] and thus adds to a considerable body of testimonial evidence from enchanted users of this technology, Eldagsen refused the award, questioning the character of the images as "photography" and calling for a discussion on how to value these AI outputs in the future.[22]

---

17 Mihály Héder, "AARON," in: Encyclopedia of Artificial Intelligence: The Past, Present, and Future of AI, ed. Philip Frana, Michael Klein, (Santa Barbara, 2021), 1f.

18 See, for example, his exhibition at MoMA titled "Unsupervised" (November 19, 2022 - April 15, 2023), where the latent space of the datascape of all digitized artworks in the museum's possession can be interactively explored, https://www.moma.org/calendar/exhibitions/5535.

19 "Edmond de Bellamy" is an ironic "translation" into French of the name of the inventor of "GANs" (General Adversarial Networks), Ian Goodfellow, the dominant generative AI architectures before the Transformer and Diffusion model revolutions; see also: Tim Schneider, "The Gray Market: How Christie's So-Called 'AI-Generated' Art Sale Proves That Records Can Distort History (and Other Insights)," Artnet, October 29, 2018, https://news.artnet.com/opinion/gray-market-obvious-portrait-1381798. The entire project was mostly seen critically; the artists used third-party code without disclosing it and made exaggerated, misleading claims of an independent AI creation, see for this criticism e.g.: James Vincent, "How Three French Students Used Borrowed Code To Put The First AI Portrait In Christie's," The Verge, Oct 23, 2018, https://www.theverge.com/2018/10/23/18013190/ai-art-portrait-auction-christies-belamy-obvious-robbie-barrat-gans.

20 See for a detailed discussion of those systems that still required skilled specialists to make them work: the chapter "Putting Computational Creativity to Work" in: Oliver Bown: Beyond the Creative Species (Cambridge (MA) 2021), 215-266.

21 Kevin Roose, "The Shift - An A.I.-Generated Picture Won an Art Prize. Artists Aren't Happy," New York Times, September 2, 2022, https://www.nytimes.com/2022/09/02/technology/ai-artificial-intelligence-artists.html.

22 However, the question seems legitimate as to why he submitted these images as "photographs" in the first place, see: N.N., "AI-Generated Art Sparks Controversy in Photography and Art Competitions," Culture.org, April 20, 2023, https://culture.org/ai-generated-art-sparks-controversy-in-photography-and-art-competitions.

At this point, it can be stated that the art world has entered an age where established professional photo jurors rate an AI-generated image spat out in a few seconds higher than anything they have seen authored by a human.

## 2.2 Text-to-image

From the user's perspective, the text-to-image process is simple: you type in some text and the computer spits out an image. The first time this process was successfully implemented was in 2014, when the image classification process (extracting keywords and captions from an image) was successfully reversed. The resulting 32 x 32 pixel color blobs with rudimentary structures were no artistic masterpieces. But they marked the beginning of a race to achieve just that: aesthetically beautiful, smart, and meaningful computer art.[23]

In the following years, a growing number of technically savvy developers and artists embraced text-to-image technology. The following series of images shows a typical example of earlier image generating technology (GANs). The resulting illustrations often lack consistency and coherence, seeming to have an expressionistic character, close to psychedelic fever dreams.

---

[23] Elman Mansimov et al., "Generating Images from Captions with Attention," International Conference on Learning Representations, ArXiv, Feb. 29, 2016, https://doi.org/10.48550/arXiv.1511.02793.

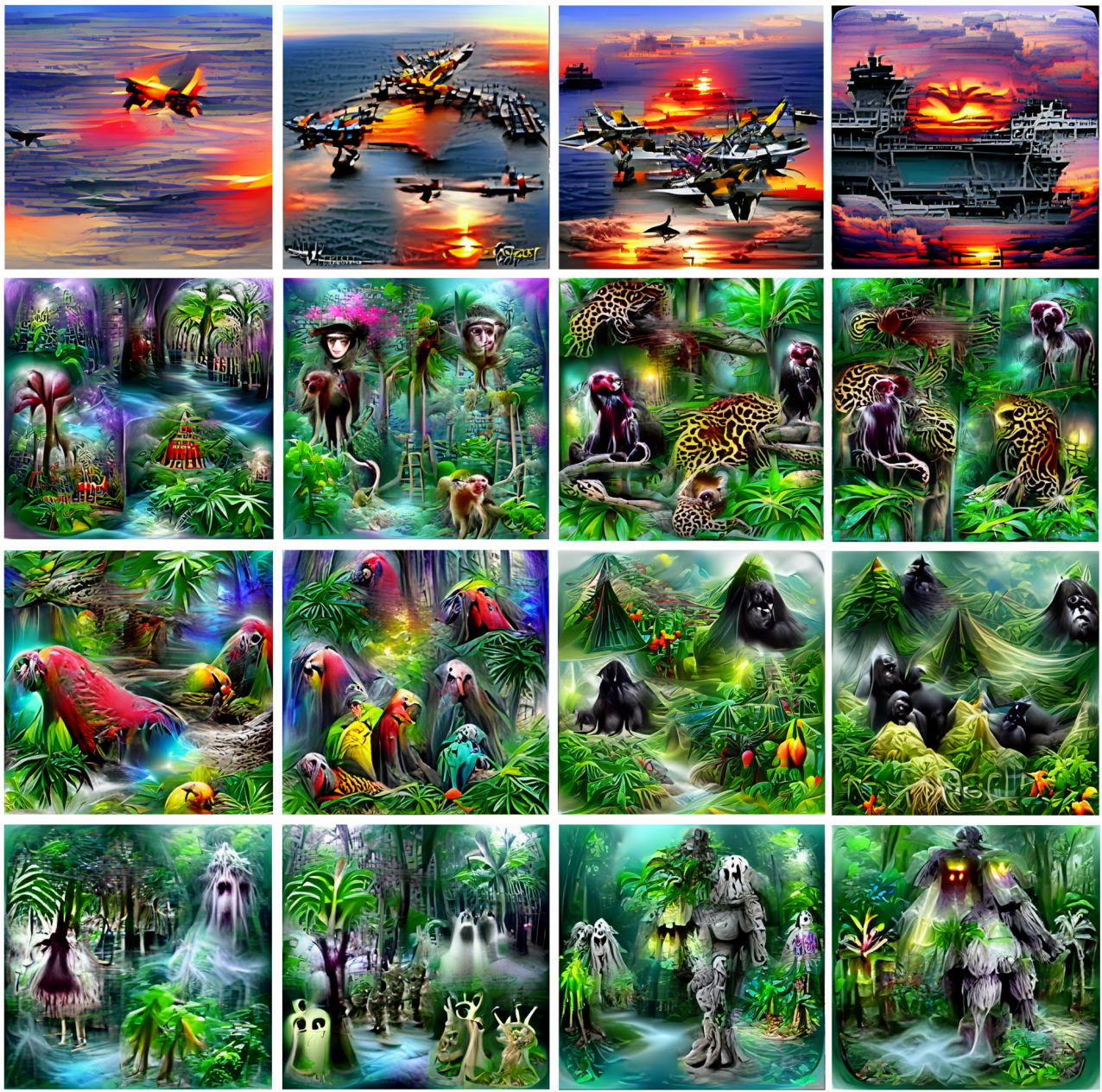

Figure 1. [Jens Knappe/ VQGAN; 42/50 Daydream, 2021]

These wild, sketchy color explosions were far from the award-winning images that would be produced a year later by image generators such as DALL E 2, Stable Diffusion and Midjourney.

**2.3 Genesis**

In spring 2022, I joined the early (highly secretive) testing phase of DALL E 2. The system was unlike anything I had seen before, exceeding the capabilities of the VGANs I had been experimenting with before. It demonstrated a deep text understanding and could execute complex requests flawlessly. The aesthetic realization quickly struck me as downright magical; the depictions were intelligent, witty, and beautiful. There was even irony in them, and I began to seriously wonder how it found its way into these images.
I have experimented with many elaborate prompts. But here I want to present an image where the text input was as simple as possible, just one word: "Genesis."

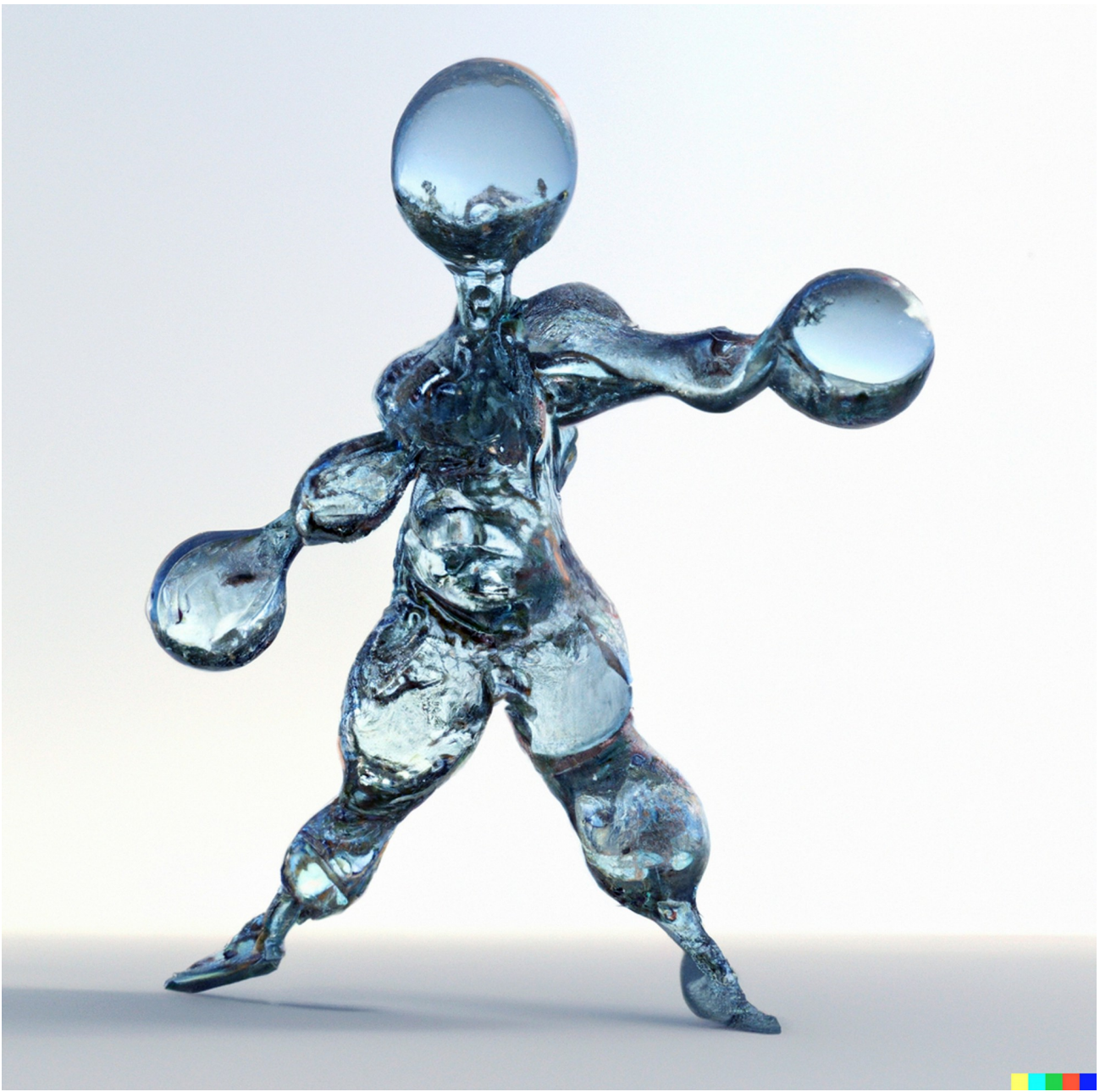

Figure 2. [Jens Knappe/ DALL E 2; Genesis, 2022]

The aim was to explore the multifaceted concept of origin, spanning biblical history to the sciences. This image shows a creature like nothing I have ever seen before in my life, made of what appears to be a large blob of water. This figure could have come into existence only recently, with the demeanor of a challenging adolescent who has just gained self-awareness and is beginning to realize its own power. One can still find the tiny creature cute, but it's obvious that there's a disturbing and unsettling element to it. Aesthetically, this image is close to perfection. There are no distractions from the character itself, it is a powerful illustration of a giant subject, the visual language is crystal clear. My part, that of the human in this collaborative effort, was almost non-existent, this artwork was created by a computer.
This image left me wondering about its source. I scoured the internet for "Genesis" illustrations and related content, but found no matches for a creature emerging from a blob of water. The closest concept with extensive imagery was the Golem myth. To this day, I've found no explanation for this AI to create such an image. As previously discussed, Machine Learning systems should operate within their training data, acting like "stochastic parrots." By design, the glass surface we see when we look at these ML AIs should be that of a mirror, never a window that opens up to true discovery and new horizons.

If a human illustrator or CGI artist had sent me this image, I would have considered it a meaningful work of art, an act of considerable human ingenuity. An abstract concept seamlessly blends with an aesthetic idea, yielding a unique, insightful, and nearly perfectly executed illustration.

**2.4 JFK – the Memoirs**

JFK at the age of 80. Everybody may have an idea of how he might have looked like, but what does that mean in terms of text-to-image? There have been AIs for some time that make an image of a person older or younger. But that is an image-to-image process, where an input picture is being changed applying aging algorithms. This picture, however, is very different. No photo of the 43 year old JFK was pasted in and an algorithm applied what it learned about aging to the image. No, this is text-to-image. The image was imagined from nothing, translated from the datasphere of text to the realm of visual information. And aside from the excellent technical quality of this picture, one thing is remarkable: There are no images of an 80-year-old JFK in the training data. This picture combines the concrete concept of a person with the abstract concept of age, creating something novel, seemingly beyond the training range.

These are just two examples from a much larger set of images that seem to challenge the idea that these systems cannot reach beyond mere statistical repetition of learned patterns.

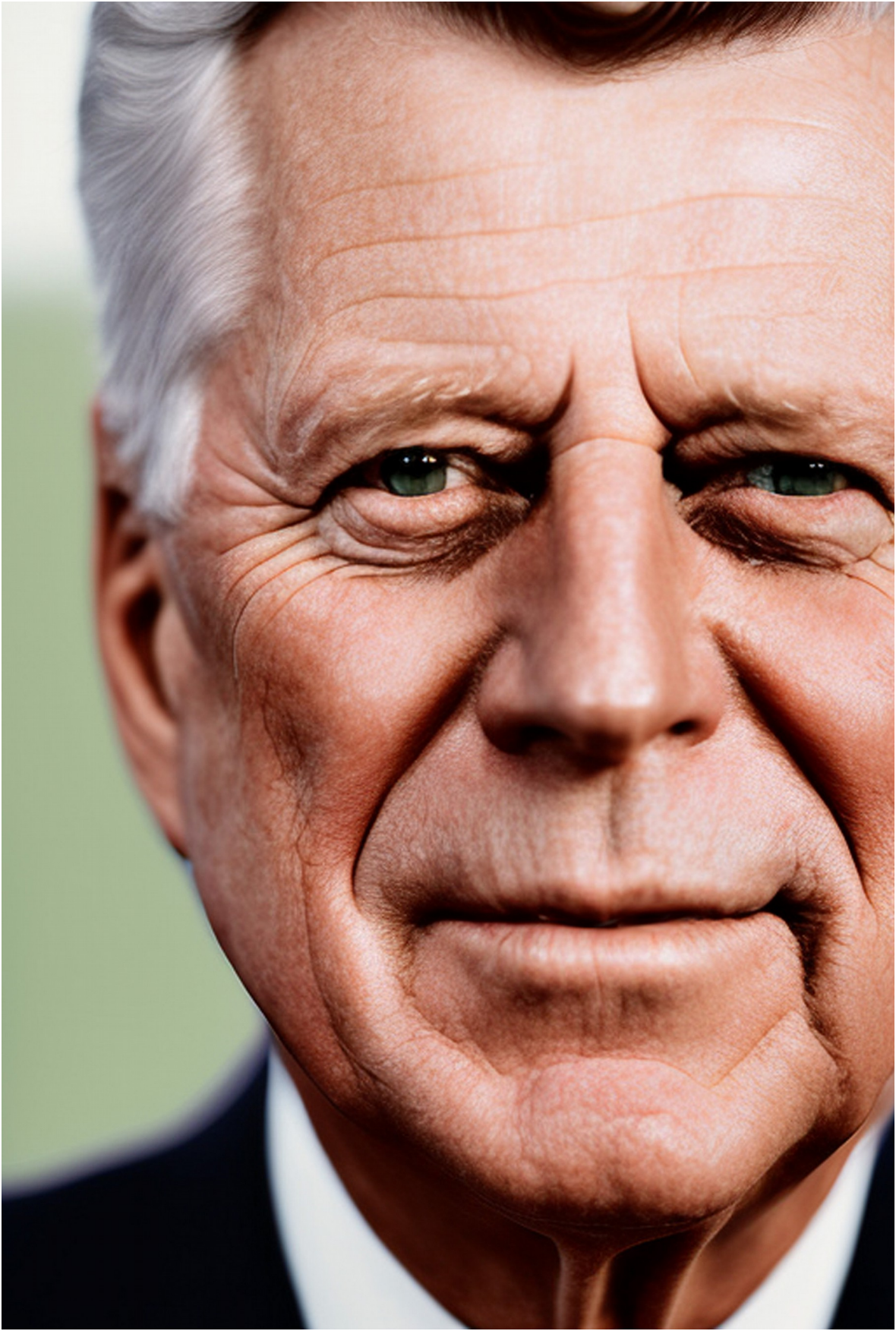

Figure 3. [Jens Knappe/Midjourney; JFK – the Memoirs, 2023]

# 3 The Technology Behind Text-To-Image

**3.1 Image Generation as Translation**

To understand the fundamental nature of the technology that is driving this qualitative shift in AI visual content creation, it is important to first get the concept right: the process of image creation does not consist of assembling and stitching together images pulled from a large database. And there is no rendering engine running anywhere on a smartphone, computer, or in the cloud, putting textures on top of threedimensional grid models, as we have with any other kind of CGI. In fact, this way of producing images in a computer is something completely new, both conceptually and practically. The best way to conceptualize the entire process of AI image creation is to think of it as an act of translation from one language to another: this time not from English to French, but from text to image, transposing semantic concepts into image outputs.[24]

Computers only "see" numbers. All text is encoded as a series of integers, as are images: they are divided into the number of pixels in an image, with each spot having exactly three color values (RGB). And the beauty of it is that these numerical representations, be it for sound, image, video, or text, "look" the same to computers and are interchangeable. And modern generative AI systems can translate and manipulate them as if from one language to another, easily tying together images with text. In order to align the embeddings - the numerical representations of text and images in the dataspace where Deep Neural Networks store their knowledge - they have acquired special similarity matrices. The most prominent of these is CLIP[25] from OpenAI.

And the translation aspect is probably the most powerful, yet most trivial technological factor that makes the generated visuals seem so genuine, original, and creative. The vocabulary of images is simply much larger than that of text: For example, an image of a cat may show more than just the isolated feline, but also include a sofa and flowers. When translating the textual prompt "cat," the corresponding vector embeddings in the image data space often retrieve a richer visual vocabulary than the original text input (with the sofa and flowers). And this surplus of visual information often leads to the impression that there is an unusually rich well of imagination on the other side.

**3.2 The Transformer Architecture**

Language comprehension and translation has been the most significant area of research in AI in recent years, and the most important technological change that has enabled today's generative systems was the invention of the Transformer architecture.[26] Transformers differ from previous architectures by prioritizing context with an "attention" mechanism, greatly enhancing "understanding" in translation and text comprehension. These are the first systems that can effectively handle fine nuances of meaning and ambiguity.

Their introduction has also accelerated technology by utilizing parallel structures in contrast to sequential tapeworm-like computations. Parallel computing is exactly how modern GPUs, graphics cards originally designed for computer games, work, and this speeds up the entire computation manyfold.

---

24 The idea of translation is well explained in: Gorti, Satya Krishna, Jeremy Ma "Text-to-Image-to-Text Translation using Cycle Consistent Adversarial Networks", ArXiv, Aug. 14, 2018, https://doi.org/10.48550/arXiv.1808.04538.
25 CLIP (Contrastive Image-Language Pretraining): method developed by the company OpenAI, in which a neural network is trained with a large set of image-text pairs. Usually, the alt-texts in web pages are used here, which contain detailed image descriptions; therefore, a very good text understanding is being achieved.
26 The Transfomer architecture was introduced by Google and described in the paper: Vaswani, Ashish, Noam Shazeer, Niki Parmar, Jakob Uszkoreit, Llion Jones, Aidan N. Gomez, Lukasz Kaiser, Illia Polosukhin "Attention Is All You Need", ArXiv, Dec. 6, 2017, https://doi.org/10.48550/arXiv.1706.03762.

Transformers are autoregressive systems, which means that a next token prediction happens here. A token can be (part of) a word or (patch of) pixel(s) in an image. The goal is a high probability next token prediction. This is the underlying mechanism most of the generative AIs we have today.[27] The autoregressive design of these architectures leads to the constant creation of tiny artifacts. At first glance this seems like an innocent little detail, but in the bigger picture it becomes the main factor in the giant leap these systems have made: The production of these infinitely small artifacts adds up to a significant act of independent creation. This phenomenon has become widely known as "hallucination," which is considered a bug in responsible AI applications, but becomes a feature in the context of creativity.

**3.3 Diffusion Models**

Diffusion Models are the technological shift integral to modern AI image generation. It involves creating visual content from complete randomness, with two key phases: the forward process destructs an image into noise, and the reverse denoising leverages learning from the forward process to create a noiseless, coherent image. Previous methods, like VAEs and GANs, had significant shortcomings, including mode collapse and struggles to create genuinely new instances, often simply reproducing training data. They generated images in a one-step process, requiring all details at the start. In contrast, the diffusion process comprises a loop of generations, improving structure and coherence with each step through probability sampling and text conditioning.[28] Diffusion models represent a significant leap in image quality and diversity. Originating from complete randomness and noise, each image is a truly unique "pixel hallucination." This results in something synthetic, novel, not similar to the training data with the produced image being just one probability that materialized out of a massive ocean of possibilities.

**3.4 Superposition: The "Creative Factors" Amplified**

None of the technologies described above were intended to make a revolutionary impact. They were all conceived as incremental improvements. At first, it wasn't even likely nor certain that they would work. Most of the progress in AI over the past decade has been driven by trial and error, and the major technological advances have often been serendipitous.[29] Each of the technologies described above represented a significant step forward in creating more coherent, unique, diverse, and original synthetic data. But in isolation, they would not have had such a profound impact on the overall process. It is the combination of these factors that makes the difference. And they do not simply add up, they amplify each other. Here we have the seed of an exponential curve. It is like an interference pattern, a phenomenon of multiple waves merging and thus superimposing. This is the same mechanism that triggers a tsunami, and it is currently at play in the development of Artificial Creativity.

---

27 In the case of text-to-image, the overall architecture is not a single Transformer as in the case of current LLMs. In the process of image generation, Transformers come into play primarily in the encoding and decoding phases.

28 For a comprehensive overview of diffusion models and technical processes, see: Ling Yang et al., "Diffusion Models: A Comprehensive Survey of Methods and Applications," ArXiv, Sept. 2, 2022, https://doi.org/10.48550/arXiv.2209.00796; for a brief discussion of the key concepts see also: Ava Amini, "MIT 6.S191: Deep Learning New Frontiers," Youtube, Jan. 12, 2023, https://www.youtube.com/watch?v=FHeCmnNe0P8. For the text conditioning aspect, see Aditya Ramesh et al., "Hierarchical Text-Conditional Image Generation with CLIP Latents," ArXiv, Apr. 13, 2022, https://doi.org/10.48550/arXiv.2204.06125, see also: Dhariwal Prafulla et al., "Diffusion Models Beat GANs on Image Synthesis," Advances in Neural Information Processing Systems (4. 2021): 8780–8794.

29 See especially for the "Transformer"-paper: Ansari, Tasmia "'Attention is All You Need' Gets No Attention - What are 'Attention is All You Need' Authors Doing Right Now – Everything About The Thinkers, Doers, and Quitters", Analytics India Magazine, February 1, 2023.

## 4 Creativity: Human and AI

### 4.1 Genius, Chaos and *Götterfunke*

It seems like a growing number of people begin to perceive generative AIs as "creativity machines," while for skeptics the term "machine creativity" remains an oxymoron. The question of creativity in machines is not banal. It rather delves into the essence of the *condition humaine*. What does it mean to be "creative" and how do modern AI systems fit into this discussion?

How new ideas enter our minds is one of the oldest puzzles of human existence. In antiquity, art was considered mere imitation of nature, while Christianity attributed new ideas to divine induction. It was not until the turn to humanism in the Renaissance that humans began to be seen as the true originators of their artistic works. The polymath "Renaissance man," a single individual who massively advanced art and science, such as Leon Battista Alberti, Giambattista della Porta, or Leonardo da Vinci, became the ideal of the universal genius. The idea of a creative individual, especially a "genius," is largely tied to Western culture. In many other parts of the world, creativity is either seen as a collective effort or they lack this concept altogether.[30] Furthermore, the "genius" myth has often formed the basis of totalitarian leadership cults.[31]

Another powerful myth suggests that great works in art, literature, science, philosophy, and music arise from a person's struggle with the world and the act of creation. Clinical studies suggest that high creativity may actually be associated with an increased likelihood of mental illness,[32] and the "Angst" of the creative act has intrigued psychoanalysis for over a century.[33] The idea of the tormented genius persists in our culture,[34] exemplified by figures like Friedrich Nietzsche, of those who "have the chaos in them in order to give birth to a dancing star."[35]

When a truly new idea arises, it is often very difficult or even impossible to reconstruct its origin. It often remains a mystery, a puzzle to its originator, and this void in conscious memory is often filled by adding a dimension to it that goes beyond everyday routine existence. This can be a trance-like state of mind or an injection from outside, from some other source which often a divine quality is attributed to.[36] Quotes from historical figures who couldn't grasp how they got their groundbreaking ideas are abundant. Rick Rubin, the renowned music producer, has written a book exploring the creative process based on his decades of experience. The book exemplifies the idea of creativity as a "*Götterfunke*," a divine spark bestowed upon a fortunate (or unfortunate) few. Rubin views artistic inspiration as "divine flash of light."[37] Artists, in his opinion, serve as simple mediums, open channels for these godly "bolts of lightning,"[38] and if they don't grab it, someone else will. It is "like the universe conspiring on our behalf, if we let it."[39] If not, it will be gone and it cannot be retrieved. But once seized, something greater than the individual(s) involved can emerge.

---

30 See articles on Asian, African and South American cultures and their concepts of creativity in: James C. Kaufman, Robert J. Sternberg (eds.), The International Handbook of Creativity (Cambridge, 2006).
31 Dictators like Hitler, Stalin and Mao had been actively portrayed as "geniuses" by the propaganda machinery of these regimes, see also: Ryan Skinnell, "The Destructive Myth of the Universal Genius," JSTOR Daily , June 14, 2023, https://daily.jstor.org/the-destructive-myth-of-the-universal-genius/, see for an early criticism by a member of the Vienna Circle: Edgar Zilsel, Die Genierelion (The Religion of Genius) (Vienna, 1918), for a critique on the male bias: Cody Delistraty, "The Myth of the Artistic Genius," The Paris Review, Jan 8, 2020, https://www.theparisreview.org/blog/2020/01/08/the-myth-of-the-artistic-genius/.
32 See James C. Kaufman, Creativity and Mental Illness (Cambridge, 2014) and Nancy Andreasen, "The Relationship Between Creativity and Mood Disorders," Dialogues in Clinical Neuroscience 10(2):251-5, Feb 2008, https://doi.org.10.31887/DCNS.2008.10.2/ncandreasen.
33 There is plenty of literature on this aspect of human creativity. One book still stands out, written by an Austrian Freud deflector: Otto Rank, Kunst und Künstler. Studien zur Genese und Entwicklung des Schaffensdranges (Gießen, 2000) (originally published in 1932 as "Art and Artist").
34 See for example: Douglas Eby, "How Pop Culture Stereotypes Impact the Self-Concept of Highly Gifted People," High Ability, https://highability.org/511/how-pop-culture-stereotypes-impact-the-self-concept-of-highly-gifted-people.
35 Friedrich Nietzsche, Thus Spoke Zarathustra (1892).

Scientific inquiry into the creative process, lately aided by neurophysiology, has yielded numerous insights, but also has fallen short in demystifying it.[40] It is also widely agreed upon that human creativity can never be a "*creatio ex nihilo*" and that there is always a lot of old in new ideas.[41] Even the most outstanding intellectual and artistic achievements of the human mind have been realized by building on earlier knowledge, by standing on the shoulders of giants. But rationalization hasn't unraveled the mystery of human creativity; in fact it has, in some ways, only reinforced all those myths. The question of creativity still retains a pre-scientific character which defies rational explanation.[42] And creativity is also widely valued as something even superior to pure intelligence, something on top of it. Something that sets us apart from anyone and anything else, as the marvel of the human mind.

When we consider the question of creativity in machines, we cannot possibly be neutral. It is as if we have spent a lifetime gazing at a brightly lit scenery and suddenly close our eyes. What remains is the glowing, gleaming, phosphorescent afterimage that seems to be etched into our retinas. And this is what we have to deal with when we think about "Artificial Creativity."

**4.2 Concepts of Creativity in Computers**

Cognitive scientist Margaret Boden has been studying (the possibility of) creativity in computers since the 1970s. [43] She argues that in principle all the ways in which humans arrive at novel ideas also stand open to artificial systems. Boden defines a creative idea or artifact as something "new, surprising, and valuable" and distinguishes three different ways of arriving at new ideas:

- Combinational: through the creation of novel combinations of familiar ideas: this could be a collage of images, analogies in science or literature;

- Exploratory: by exploring the potential of conceptual spaces.[44] Boden sees this type of creativity as the way most artworks materialize, and how adventurous these explorations are determines the degree of creativity;

- Transformational: by making transformations of the conceptual space itself that enable previously impossible ideas to emerge.[45] This is obviously the most ambitious and difficult way to come up with something novel.

---

36 A historical example would be that of Johann Wolfgang Goethe, who testified that he wrote his first novel ("Die Leiden des jungen Werther") "almost unconsciously, like a somnabulist" (Peter de Kuster, "How Goethe Became Notoriously Famous at 25," The Hero's Journey Today, June 7, 2020, https://theherojourneytoday.wordpress.com/2020/06/07/how-goethe-became-notoriously-famous-at-25/) and for the impression of divine inception, Jack Kerouac, who seriously thought his novel "On The Road" was "dictated by the "Holy Ghost": Katharine Webster, "Happy Birthday, Jack Kerouac!" UMass Lowell, Jan 19, 2021, https://www.uml.edu/news/stories/2022/kerouac-centennial.aspx).
37 Rick Rubin, The Creative Act. A Way of Being (New York, 2023), 127.
38 Ibid, 289.
39 This quote is from a conversation Rubin had on the "Daily Stoic" Youtube channel: Daily Stoic, "Rick Rubin on The Creative Act, Overcoming Ego, and Enjoying the Process," Youtube, June 9, 2023, https://www.youtube.com/watch?v=Zr-OZaJ-DVQ.
40 See for an overview of current theories e.g.: James C. Kaufman (Ed): Cambridge Handbook of Creativity (Cambridge, 2019).
41 Alan Turing, "Computing Machinery and Intelligence," Mind, vol. 49 (1950): 433-460; see also Thomas Ward, "What's Old About New Ideas," in: The Creative Cognition Approach, ed. Steven Smith et al. (London, 1995), 157–178, https://doi.org/10.48550/arXiv.2204.06125.
42 As John Searle put it in reference to the "hard problem" of human consciousness: John Searle, "How to Study Consciousness Scientifically," Brain Research Reviews, 26/ 1998: 379-387.
43 Margaret A.Boden, The Creative Mind: Myths and Mechanisms (London, 2004).

Boden sees all three paths open to AI. However, she wrote most of this while the leading paradigm in AI was that of symbolic representation. In essence, this means hard-coding (logical) rules into the computer. And of course there is nothing in principle against programming a way to transform conceptual spaces, to go beyond a given dataset. The issue of transformational creativity in current ML architectures will be discussed in more detail later.

**4.3 Interpolation, Extrapolation, and Out-of-the-Box Thinking**

Demis Hassabis, head of "Google Deepmind,"[46] examines creativity in these ML architectures using computer science terminology. He identifies three levels: On the one hand, there are systems that are able to interpolate, i.e. find statistical commonalities about the known and thus elementary laws. This "low-level" form of creativity is basically the "DNA" of most ML systems.[47] While this down-to-earth basic creativity is fulfilled in these systems by design, it gets more intriguing when an AI is able to draw conclusions about the unknown from the known. Hassabis terms this second level of creativity as "extrapolation." This idea is close to Boden's concept of exploratory creativity and already contains transformational elements. Hassabis believes this is rarely achieved in artificial systems, an example being Deepmind's AI "AlphaGo" making the unprecedented, game-winning "move 37" in the Go game against former world champion Lee Sedol in 2016, breaking with centuries-old conventions of the game.[48]

Hassabis defines the highest level of creativity as "out-of-the-box thinking" or simply "invention." This would be, for example, an AI inventing a game like chess or Go, or coming up with something like Einstein's theory of relativity. This represents the pinnacle of creativity, akin to pure "genius,"

---

44 Boden defines those as "structured styles of thought", see: Margaret A. Boden, "Creativity in a Nutshell," Think, 15/2007: 83-96, https://doi.org/10.1017/s147717560000230x.
45 Margaret A. Boden, "Creativity and Artificial Intelligence," Artificial Intelligence, 103/1998: 347-354.
46 Hassabis, who was recently awarded the Nobel Prize for the AlphaFold system, has a degree in computer science and a Ph.D. in cognitive neuroscience. His company, Deepmind, which aims to create AGI, was acquired by Google in 2014 but has remained largely independent as a research-focused unit headquartered in London. However, in May 2023, given the growing commercial importance of AI applications to Google, it was officially merged with the company's other AI division, "Google Brain," renamed "Google Deepmind," and moved to Mountain View.
47 In creative practice, this means that these systems are good at finding analogies.
48 For the concepts of AI creativity: Demis Hassabis, "DeepMind: From Games to Scientific Discovery," Research-Technology Management, 2021: 18-23, https://doi.org/ 10.1080/08956308.2021.1972390 or as a lecture: Demis Hassabis, "Creativity and AI – the Rothschild Foundation Lecture," YouTube, https://www.youtube.com/watch?v=d-bvsJWmqlc; for a discussion of "Move 37" see Marcus Du Sautoy, Creativity Code: How AI is Learning to Write, Paint and Think (Cambridge (MA), 2019) and Jens Knappe, Genesis. A Creation Story in Cooperation with an Artificial Intelligence (Berlin, 2022), 32-34.

and achieving this would pave the way to a radically superior artificial intellect, a "super-intelligence."

## 4.4 Artificial Creativity

The issue of Artificial Creativity[49] was long seen as essentially theoretical, with symbolic models considered merely mimicking creative processes, copied from human behavior and transferred into a large bowl of spaghetti code; in principle nothing more than a barrel organ playing a symphony. This changed with the advent of Machine Learning content generation systems in the 2010s, especially with GANs.[50] It struck many that the artifacts produced by these systems, all those psychedelic images (like the pictures of the "Daydream" series), would have qualified as proper works of art if they had come from a person. This led to profound curiosity about the inner workings of these Deep Neural Networks. After all, these architectures were modeled after the biological brain, the only use case we know of in the entire universe that has been proven to produce true creativity.

A scientific debate on the true nature of outputs from generative ML systems emerged and gained traction. A definition of Artificial Creativity in Machine Learning, still relevant today, is "the philosophy, science and engineering of computational systems which, by taking on particular responsibilities, exhibit behaviors that unbiased observers would deem to be creative."[51] Two words are important here: "responsibility" and "unbiased." These systems should possess independence, not just serving as auxiliary tools, and their products should be judged blindfolded.[52]

The most comprehensive study of Artificial Creativity has covered ML systems, including current LLMs like GPT3, assessing their creative potential in terms of Boden's criteria.[53] The authors examine 7 AIs, attributing exploratory creativity to all, combinational to 3, and the highest degree, transformational creativity, also to 3.[54] However, upon closer examination, these evaluations appear somewhat arbitrary. It is also interesting to note that while Transformer architectures are credited with the potential for transformational creativity in this study, the very same authors categorically deny this to current LLMs (which are all Transformers) a year later due to their "inner autoregressive nature."[55] Another review of 31 creativity case studies in ML systems reveals a heterogeneous pattern, with all categories detected in their outputs.[56] In a recent study, ChatGPT outperformed 99% of humans on a formalized test of creative thinking.[57] Arthur I. Miller, who studied human creativity in innovators and geniuses like Einstein and Picasso, concludes that ML systems inherently possess all the qualities that outstanding people have, in his unique categorization these are 7 high creativity traits and 2 for "genius".[58]

---

49 In this text, the term "Artificial Creativity" is used as a proper name. However, "computational creativity" seems to be the term most often used in an academic context, followed by "machine creativity." To name this phenomenon can also be found: artificial imagination, creative computing, mechanical creativity, creative AI. I prefer the term "Artificial Creativity" with capital letters for two reasons: First, the most common terms are, in my opinion, too specific to today's silicon-based computers. Currently, there is great progress in the work with biological, carbon-based structures (like the neurons having achieved mastery in the Pong game by Cortical Labs) and this bionic technology, neuromorphic computing, might play a more important role soon. The most important reason however is, that the term AC implies a value similar to that of AI itself while the others imply to be a mere subfield of AI. I think both go hand in hand and the question of creativity is of crucial importance in achieving anything like AGI or on the path towards "superintelligence." However, for the sake of play with words, I also use the term "machine creativity" in this article.

50 In the field of visual arts and illustration. At the same time, systems that produced music and poetry also evoked this kind of interest in their creative potential.

51 Simon Colton et al., "Computational Creativity: The Final Frontier?" 20th European Conference on Artificial Intelligence (Montpellier), Vol. 242/2012, 21–26.

52 Ibid.

53 Giorgio Franceschelli et al., "Creativity and Machine Learning: A Survey," ArXiv, July 5, 2022, https://doi.org/10.48550/arXiv.2104.02726. The survey also provides an overview of the most promising formalized tests for evaluating creativity in artifacts. For a comprehensive overview of formalized evaluation methods, see also: Sameh Said Metwaly et al., "Approaches to Measuring Creativity: A Systematic Literature Review," Creativity. Theories - Research - Applications, 2/2017, 238–275.

54 Franceschelli et. al 2022, 16.

It is evident in all the studies that there is a problem with the robust evaluation of creativity, even with formalized rating methods. Assessing creative value remains a matter of human, subjective judgment. Interestingly, in dozens of these studies, not a single one denied creative capabilities to any of the ML systems investigated. Surprisingly, even the most advanced form of creativity, the transformational category, was found in many systems. Yet, as outlined above, going beyond the boundaries of the training datascape and the so derived operational framework of conceptional space seems to involve some kind of Voodoo element: the plain textbook answer would be that this should not be feasible in an (autoregressive) ML architecture.

## 4.5 Transformational Creativity – the Vastness and Massive Multidimensionality of Conceptual Space

The question of the transformational quality seems to be today's frontier of Artificial Creativity. At its core lies the impression many of those testing and using these systems have that the output of these AIs is more than the sum of its parts, venturing into uncharted territory and thereby creating something novel, surprising and valuable. This issue delves deep into Neural Networks' nature, offering potential insights to demystify the black box.

There is one common misconception: that the original training data is in any way part of the ML system, be it in a database or stored anywhere else. This is not the case. The data is used for learning and extracting statistical insights, creating a blueprint for construction, akin to biological DNA. ML systems encode learned features in high-dimensional vectors, exceeding human limitations to three dimensions. As systems advance, "feature vector" spaces can encompass hundreds or even thousands of dimensions in image generators.

These spaces are not profane. The sheer vastness of their geometry harbors the potential for serious serendipity. Stephen Wolfram has scrutinized the open source image generator Stable Diffusion.[59] He identifies more than two thousand dimensions in the feature space of this system. These spaces contain numerous "islands" of semantic meaning within a vast "interconcept space," of images for which we have no words, no mental "concepts." Images in this wide open space are often "weird, sometimes disturbing," they often look "bizarrely magic to us." Often they could very well pass for "mindful interpretations."[60] Wolfram gives the small spots of "meaning" a concrete volume: they occupy about $10^{-600}$ of the total space. This is a number with 600 zeros, known as centillion, dwarfing the estimated number of atoms in the observable universe (between $10^{78}$ and $10^{82}$.)[61]

---

55 Giorgio Franceschelli et al., "On the Creativity of Large Language Models," ArXiv, July 9, 2023, 8, https://doi.org/10.48550/arXiv.2304.00008.

56 In a review of 31 papers they found computational creativity everywhere and also all categories of creativity in pre-Transformer ML systems: 16 combinational; 21 exploratory; transformational: 24 search space transformation and 13 boundary transformation: Marcus Basalla et al., "Creativity of Deep Learning: Conceptualization and Assessment," Proceedings of the 14th International Conference on Agents and Artificial Intelligence (ICAART 2022), Vol. 2, 99-109, 104, DOI: 10.5220/0010783500003116.

57 Cary Shimek, "UM Research: AI Tests into Top 1% For Original Creative Thinking," UM News Service, July 5, 2023, https://www.umt.edu/news/2023/07/070523test.php. The "Torrence Test of Original Thinking (TTCT)" was used here. See for further discussions of human evaluations of human vs. computer artworks: Martin Ragot et al., "AI-generated vs. Human Artworks. A Perception Bias Towards Artificial Intelligence?" CHI EA '20: Extended Abstracts of the 2020 CHI Conference on Human Factors in Computing Systems, April 2020, 1–10, https://doi.org/10.1145/3334480.3382892 and Joo-Wha Hong, "Bias in Perception of Art Produced by Artificial Intelligence," in: Human-Computer Interaction. Interaction in Context, ed. Masaaki Kurosu (Cham, 2018), 290-303.

58 Arthur I. Miller, The Artist in the Machine (Cambridge (MA), 2019), 307-309.

Some fundamental properties of mathematics seem to collapse under the sheer weight of this dimensional toomuchness. In spaces over 100 dimensions, "interpolation is doomed by the curse of dimensionality,"[62] as LeCun et al. point out. This basically means that any mathematical operation within such a fragile conceptual meaning space inevitably causes a result outside of it, in the weird vastness of what Wolfram referred to as "interconcept space." And Bubeck, the author of the study on emergent phenomena in raw GPT4, points out: "Beware of trillion-dimensional space and its surprises."[63] This relates to the vast number of adjustable parameters the latest LLMs have.[64] Conceptual spaces have simply grown beyond human comprehension. And so results that seem to be transformational may also very well lie inside the original conceptual space.

While not conclusive, these findings suggest that the nature of the spaces in which ML AIs acquire knowledge may contribute to the impression of results transcending the original training data, of transformational creativity where it shouldn't be. Be it a tiny figure apparently made out of a blob of water illustrating a "Genesis" myth or a portrait of the 80-year-old JFK.

**4.6 Hybrid Systems as a Way to Achieve Transformational, Out-Of-The-Box Creativity**

LLMs and text-to-image systems have grown in sophistication internally but retained simplicity in their basic architecture as ML systems. There's an ongoing debate about the limits of Transformer architectures, their shortcomings, and the quest to build improved AIs that may become the form factor for AGI. Current architectures have "smooth scaling laws" going for them, which means that up to this point, the more compute is put in, the more (predictably) powerful these systems get.[65] Some, like Chomsky and Marcus, view these architectures as an error-prone dead end with unhedgeable hallucination issues,[66] while LeCun anticipates Transformers becoming outdated soon. He suggests a redesign towards "objective-driven AI" with a "modular cognitive architecture" for better reliability,[67] as he sees Transformers only as a rather inefficient and unreliable transitory technology.

---

59 Stephen Wolfram at a conference on ChatGPT and Mechanistic Interpretability at the MIT Department of Physics, also documented online: https://www.youtube.com/watch?v=u4CRHtjyHTI.; see also: Stephen Wolfram, "Generative AI Space and the Mental Imagery of Alien Minds," Stephen Wolfram Writings, 2023, https://writings.stephenwolfram.com/2023/07/generative-ai-space-and-the-mental-imagery-of-alien-minds.
60 Notes taken from the conference talk.
61 This number is so enormous because the volume of an object in a space decreases with each additional dimension, while the size of the entire space increases. This can be easily visualized by a circle in a given 2-dimensional space becoming a sphere in 3 dimensions and becoming smaller relative to its surroundings. This space would become exponentially smaller with each additional spatial dimension.
62 Randall Balestriero et al., "Learning in High Dimension Always Amounts to Extrapolation," ArXiv, Oct 29, 2021, 2, https://doi.org/10.48550/arXiv.2110.09485.
63 Brian Wang, "Sébastien Makes the Start of AGI Case for GPT-4," Next Big Future, April 8, 2023, https://www.nextbigfuture.com/2023/04/sebastien-makes-the-start-of-agi-case-for-gpt-4.html.
64 These are the weighted connections between the (simulated) neurons in the model. The analogy in the human brain would be a synapse between neurons.
65 The most prominent proponents of Transformers come from the OpenAI-Microsoft universe, most notably (former) OpenAI chief scientist Ilya Sutskever, who believes that the shortcomings of Transformers will ease the larger these models become, see: Dwarkesh Patel, "Ilya Sutskever (OpenAI Chief Scientist) - Building AGI, Alignment, Spies, Microsoft, & Enlightenment," Youtube, March 27, 2023, https://www.youtube.com/watch?v=Yf1o0TQzry8. See also: Reid Hoffman, "Sam Altman and Greg Brockman on AI and the Future (Full Audio)," Youtube, Aug 2023, https://www.youtube.com/watch?v=rd1pwJJNNr8.

A crucial question in this discussion is implementing something like a grounding in the world, some kind of "common sense" and "alignment" with human values in powerful AIs. In order to give these systems an objective function, some kind of trajectory, an old idea may see its resurrection: symbolic representational models.

And this combination, "hybrid machines"[68] with "meta-algorithms"[69] also has the potential to significantly expand transformational creativity in these systems. A meta-algorithm may guide a ML system into new territories. It might supercharge these models by systematically transforming the search space as well as the boundary of the conceptual space itself.[70] AlphaGo, the very AI that produced the miraculous "move 37," has elements of the best of both worlds in its architecture: a so-called Monte Carlo tree-search algorithm directs the ML system in the most promising direction to put its brute deep neural force to work.[71] Hybrid systems like these may standardize transformational and even tapping into inventory, out-of-the-box creativity.

## 5 Art World: The Question of Authorship

**5.1 Artificial Creativity in the World of (Visual) Art**

So far, the technological potential for Artificial Creativity and AI-generated artworks has been examined in isolation. However, the situation with art can obviously be much more complex than this: artworks are contextualized in an entire cosmos of social actors like the artist, general and expert audiences, critics and peers as well as historically. Other factors, like the production process itself as well as the context with other artworks, also play a role in the assessment of artistic creativity.[72]

Discussing this issue means delving deeper into the "value" dimension. What makes an artifact produced by a system employing Artificial Creativity valuable in the context of (visual) art? This area lacks clear metrics, and much of what follows is speculative in nature. There are a number of unknowns, obviously also due to the fact that AI art today – despite everything described so far - has only materialized into an embryonic state and much of what is to come is highly uncertain at this point.

---

[66] See Noam Chomsky, "The False Promise of ChatGPT," The New York Times, March 8 2023, https://www.nytimes.com/2023/03/08/opinion/noam-chomsky-chatgpt-ai.html and: "Web Summit: Debunking the great AI lie: Noam Chomsky, Gary Marcus, Jeremy Kahn," Youtube, Nov 14, 2022, https://www.youtube.com/watch?v=PBdZi_JtV4c.

[67] Yann LeCun at the MIT conference "The Impact of ChatGPT" of 2023; see also the recording of it online: https://www.youtube.com/watch?v=vyqXLJsmsrk. See also: Yann Lecun, "A Path Towards Autonomous Machine Intelligence," Open Review, June 27, 2022, https://openreview.net/pdf?id=BZ5a1r-kVsf.

[68] Miller, 307.

[69] Du Sautoy, 279.

[70] In transformational creativity, boundary transformation has a higher potential to lead to a paradigm shift, comp. Basalla et.al,. see also: Geraint Wiggins, "A Preliminary Framework for Description, Analysis and Comparison of Creative Systems," Knowledge-Based Systems, 19/2006, 449–458.

[71] See for AlphaGo's architecture the description on the Deepmind website: https://www.deepmind.com/research/highlighted-research/alphago.

[72] See also: Ramón López de Mántaras, "Artificial Intelligence and the Arts: Toward Computational Creativity," Open Mind BBVA: The Next Step: Exponential Life (Madrid, 2017), 99-123.

In order to get the full picture, to achieve a comprehensive analysis of Artificial Creativity in the context of (visual) art, the issues of authorship and the art world will be discussed.[73]

**5.2 Agency and Authorship: Tool → Collaborator → Independent Agent**

Image-generating capabilities can be systematically applied to find, as AI artist Sarah Meyohas puts it, "images one couldn't have imagined"[74] and it seems inevitable that AI art will come closer to the ideal of creating something "so beautiful that nobody can bear it."[75] Others predict a future "where anyone can write at the level of the best writers, paint like the great masters, and even discover new forms of creative expression."[76] However, this comes at a "cost": allowing generative AI more autonomy, granting it more "responsibility." Currently, it seems that these AIs have to be elevated from tools to true artistic collaborators.[77] The subsequent stage – which still seems a long way off – would be that of an independent agent.[78]

Collaborators and independent agents translate into AI-powered artist personas. Since the release of Chat GPT, there has been a rush for "AI agents."[79] To date, AI agents have appeared in almost every area of white-collar jobs, from ordinary office work to software engineering to, most recently, fully automated synthetic scientists. These agents conduct independent research, write papers, peer-review and publish them, and implement the scientific knowledge obtained in further research.[80]

---

73 Thereby, all aspects of the still most widely used framework for analyzing creativity in social settings, the so-called 4P model, which was developed by Rhodes six decades ago, are covered. The 4 Ps are: Person/Producer, Product, Process, and Press/Environment, see: James Melvin Rhodes, "An Analysis of Creativity," Phi Delta Kappan, Vol. 42/ 1961, 305–311, see also in respect to Artificial Creativity: Anna Jordanous, "Four PPPPerspectives on Computational Creativity," Theory and in Practice. Connection Science, 28,2/2016, 194–216.
74 Anne Ploin et al., AI and the Arts. How Machine Learning Is Changing Artistic Work (Oxford, 2022), 42.
75 This is a quote from the Chinese sculptor and performance artist, Zhang Huan; the full quote reads:"I want it to be 70 per cent beautiful, 15 per cent surrealistically beautiful, and the rest so beautiful that nobody can bear it": Rupert Christiansen, "Zhang Huan: From Baroque to Beijing," The Telegraph, Aug. 21, 2009, https://www.telegraph.co.uk/culture/art/art-features/6043643/Zhang-Huan-from-baroque-to-Beijing.html.
76 de Mántaras, 119.
77 The aspect of what is called "cooperation" or "collaboration" in this article is at the heart of the interest in artists' interaction with AI. Obviously, (almost) any use of AI systems in an artistic context involves such an element. There is a wide range of possible degrees of collaboration, depending on how much autonomy and responsibility the AI system in question actually has. This aspect has attracted some research interest, and there is some literature on the subject that also uses terms such as "assisted creation" or "co-creation." See for a discussion of this aspect: Eva Cetinic et al., "Understanding and Creating Art with AI: Review and Outlook," ArXiv, Feb 18, 2021, https://doi.org/10.48550/arXiv.2102.09109, 9f.; for a discussion on a co-creative agent framework: Matthew Guzdial et al., "An Interaction Framework for Studying Co-Creative AI," ArXiv, March 22, 2021, https://doi.org/10.48550/arXiv.1903.09709. For an example of "co-creation" in sketching, see: Devi Parikh et al., "Exploring Crowd Co-creation Scenarios for Sketches," ArXiv, May 22, 2020, https://doi.org/10.48550/arXiv.2005.07328.
78 For the introduction of the concept of "creative autonomy" of an AI system, see: Kyle E. Jennings, "Developing Creativity: Artificial Barriers in Artificial Intelligence," Minds and Machines, 20/4/2010, 489–501.
79 From early demonstrations in 2023, such as "Baby AGI" or "Chaos GPT," to OpenAI's announcement of a marketplace for such agents, called "GPTS," in November 23, to widespread adoption in 2024.
80 Chris Lu, Cong Lu, Robert Tjarko Lange, Jakob Foerster, Jeff Clune, David Ha, "The AI Scientist: Towards Fully Automated Open-Ended Scientific Discovery," arXiv, Aug 12, 2024, https://www.arxiv.org/pdf/2408.06292.

Currently, "agency" in these systems is applied by setting up routines on top of LLMs. So right now the agent is not within the core AI architecture. But there are major efforts to change this, advancing from Transformers to systems that contain an understanding of the world and develop agentic properties from within. One way or another, the "AI-powered artist persona" comes within reach.

In the art world, the issue of synthetic authorship directly challenges some of the core beliefs about human creativity and its valuation as outlined before. And even to a more general public, the idea of serious agency in AI seems far-fetched at this point. However, this may change.

**5.3 Digital Twins, Counterfeit People, Constructs**

The mass exposure to LLM-driven chatbots in general, but especially the idea of "digital twins," may be a decisive factor in pulling a substantial number of people into the gravitational field of this orbit of ideas. In a nutshell, the story goes like this: Today, there are numerous LLMs available, both proprietary and open source, which can be fine-tuned with relative ease. This can be done in order to make a perfect "personal assistant,"[81] a system that soaks up everything its host does, says, writes, and thinks. After a short while, an AI like that will be able to convincingly mirror elements of the character of its host person.[82] In combination with already sophisticated methods of voice-cloning and lifelike avatars,[83] the result could be something like a "digital twin,"[84] or - highlighting potential risks - "counterfeit people", as the late Daniel Dennett put it.[85] While these entities lack embodiment and undisclosed memories, they offer the broad intellectual scope of an LLM and 24/7 availability. These twins may become the doorkeeper and storefront many people will get used to communicate with before getting to the real person behind it.

---

81 Or "personal intelligence" (PI). That's what Inflection AI, a startup founded in early 2023 by former Deepmind co-founder Mustafa Suleyman, named this. The company later became part of Microsoft, and Suleman is now the head of Microsoft's AI division. See: Alex Konrad, "Inflection AI, Startup From Ex-DeepMind Leaders, Launches Pi — A Chattier Chatbot," Forbes, May 2, 2023, https://www.forbes.com/sites/alexkonrad/2023/05/02/inflection-ai-ex-deepmind-launches-pi-chatbot/. Another company that has been working on AI-powered conversation partners since 2016 is "Replika." Their most successful product is the (controversial) "romantic partner" mode, which allows premium users to engage in sexually explicit chats with a 3D avatar of their choice. When this feature was temporarily disabled in February 2023, it left many users with serious heartache and utter despair, see also: N.N., "Hot Bot Turns Cold Why Replika's Chatbot Stopped Flirting With Users," Deep Learning AI, March 1 2023, https://www.deeplearning.ai/the-batch/why-replika-chatbot-stopped-flirting-with-users/?ref=dl-staging-website.ghost.io.
82 The most prominent example of this in the academic world is the "DigiDan" version of the philosopher Daniel Dennett, where the GPT3 model was finetuned with some of Dennett's work, and the resulting "DigiDan" clone produced statements that were indistinguishable from those of the real Dennett; the paper also gives a very accessible description of how finetuning a general model towards a clone of a person can be facilitated: Eric Schwitzgebel et al., "Creating a Large Language Model of a Philosopher," ArXiv, May 9, 2023, https://doi.org/10.48550/arXiv.2302.01339.
83 There are many examples of this online. Very convincing is that of tech investor and enthusiast Peter H. Diamandis, who had an LLM finetuned with his own publications, as well as an avatar and a speech model taken and talks to his resulting "digital twin": Peter H. Diamandis, "A Conversation With My AI Clone on the Future of AI | EP #62," Youtube, Sept 8, 2023, https://www.youtube.com/watch?v=60StHg1eOuM. This already comes pretty close to the "doorkeeper" version described. Another example of how easy it is to create such a synthetic twin is that of Youtuber MattVidPro, who talks to an avatar created from just a few photos of him and a dozen spoken sentences, see: MattVidPro AI, "I Spoke to my AI Clone - What He Said Shocked Me," Youtube, Feb 8, 2023, https://www.youtube.com/watch?v=vwuiEJJCOZ8.
84 This idea is considered one of the most commercially viable in AI as a consumer product, and major companies are working on it seriously. Currently, the term "digital twin" is mostly used in the context of the "metaverse" to describe real-world objects, such as entire factories that are completely digitized to simulate work processes, etc. However, the same term is used to denote the idea of AI-powered replicas of real people. Lately, also the term "AI clone" has become popular. See also: Cordis Research, "Meet Your Digital Twin, a Virtual Version of Yourself. Our Thinking Digital Twins are on Their Way," European Commission Cordis, July 28, 2022, https://cordis.europa.eu/article/id/441913-trending-science-meet-your-digital-twin-a-virtual-version-of-yourself, and also: Anthony Green, "How AI is Helping Birth Digital Humans That Look and Sound Just Like Us," MIT Technology Review, Sept 29, 2022, https://www.technologyreview.com/2022/09/29/1060425/seeing-double-ai-births-digital-humans/.

There might emerge a plethora of compelling AI-powered replicas of real people: Many will create 'digital twins' of themselves and family members, and may be willing to pay to confer with those of their idols.[86] These entities may become cherished companions, friends, or sources of consolation, even allowing conversations with deceased loved ones, their virtual resurrection.[87]

Digital twins may increasingly be seen as true digital minds. Many will gain the impression that these units have consciousness and are sentient in themselves. That they are independent characters. People will start to think there is a precious soul hidden somewhere in the impenetrable fortress of a cold supercomputer cluster datacenter stuffed with thousands of NVIDIA Blackwell GPUs. This may raise serious ethical and philosophical questions. If such embeddings in a vectorized conceptual dataspace are the only remnant of a deceased loved one and they get accidentally erased by a service provider, does this amount to involuntary manslaughter? Models may be merged or take hybrid forms, possibly combining multiple celebrities or popular AI models.[88] The idea of the "essence" of a human might come up, which could be captured to various degrees in such an entity. We might be on the verge of a reality full of what William Gibson called "constructs" in Neuromancer without being prepared for it.

All this, which sounds like a distant science fiction scenario but may materialize soon, might make many people take AI agency seriously. And this may also be the trajectory on which the AI-powered artist persona may pop into existence. However, if all this does materialize, one question remains: what might the true colors of synthetic agency be?

**5.4 AI Agency and Alien Minds: The Inner Motivation to Create in a Machine**

Even with the facilitation and widespread acceptance of synthetic creative agency, the - philosophical - question remains as to how deep, authentic, and true the inner motivation, drive, and desire to create of such an AI can be.

---

85 Daniel Dennett, "The Problem With Counterfeit People," The Atlantic, May 16, 2023, https://www.theatlantic.com/technology/archive/2023/05/problem-counterfeit-people/674075/.

86 An early example is that of Snapchat influencer Caryn Marjorie who had created such an AI clone of herself that can engage in "erotic discourse" and discuss sexual scenarios with her fans who have to pay for this. This virtual twin is financially very successful and there is a waiting list for premium membership with all "sexual features", see: Lauren Haughey, "How Much Would YOU Pay For a Virtual Girlfriend? Influencer Hires Out an AI Version of Herself To Men – For a Hefty Price," Daily Mail, May 11, 2023, https://www.dailymail.co.uk/sciencetech/article-12071693/Virtual-girlfriend-Influencer-charges-men-date-AI-version-herself.html.

87 There are already numerous examples of "virtual resurrections": These are not AI-powered characters, but "traditional" CGI avatars, and when put on stage a modern version of Pepper's Ghost technology is used. Examples of such virtually resurrected celebrities are: Laurence Olivier, Gene Kelly, Marlon Brando, Marlene Dietrich, Marilyn Monroe, Grace Kelly, Brandon Lee, Steve McQueen, Bruce Lee, Audrey Hepburn, Michael Jackson. The rapper Tupac Shakur became more successful as a virtual resurrection than he was until his early violent death in 1996. Today, his "Pepper's Ghost" projection headlines major festivals like "Cochella" and fills stadium-sized venues.

88 There are already numerous examples of synthetic AI characters that have become popular celebrities in their own right; the most recent one is the AI-driven digital pop star "Noonoouri", who was the first synthetic character to be officially signed by a major record label, see: N.N., "Warner Music Sign First Digital Character Noonoouri and Release Debut Single," The Argus, Sept. 1, 2023, https://www.theargus.co.uk/leisure/national/23762886.warner-music-sign-first-digital-character-noonoouri-release-debut-single. A similar synthetic (manga) pop star that has been around since 2007 is "Hatsune Miku".

Since generative AI has become a popular phenomenon in 2022, skeptics could be heard saying something like this: true AI will not be there when it writes poetry (which it already does very well), but when it wants to write it. This meme mantra seems to represent the last line of defense against Artificial Creativity, while it seems to have become clear that all the terrain up to this point has been irrevocably lost. And this moat appears insurmountable, since it boils down to one of the most fundamental questions of human existence, one to which no answer has been found despite a several millennia lasting quest for doing so, the "hard problem of consciousness."[89] What the true nature of this inner experience, our howitistobelikeness is, whether it is purely physical or something else entirely, is as obscure today as it was in the days of Aristotle. The hard problem has not softened a bit. And there is also no scientific method in reach which could detect "real" consciousness and sentience in others - humans, animals or machines.

Current and future AIs may increasingly be perceived as conscious beings. As outlined before, this issue has already been debated in regard to current LLMs[90] and will most probably be discussed with increasing seriousness in the future. If a sophisticated AI that can pass something like the Turing Test,[91] "claims" that it wants to write poetry, this should be accepted as if a human said so. The question: "Is it really real?" should be avoided because there is no prospect for a valid answer (and most probably never will be).

We tend to equate our inner experience with consciousness as such. But it should be kept in mind that, as Anil Seth puts it, "we just inhabit one small region of a vast space of possible minds."[92] If they really emerge, these alien AI minds would likely be structured very differently. And when such an entity claims that it wants to write poetry, one should be aware that - in case it originated in a true conscious experience - this might differ fundamentally from human motivation. We would simply not be able to equate our inner experience with that of such an AI.[93] What lies beyond might be much weirder, much more alien than anything anyone could possibly imagine.

---

[89] Comp. David Chalmers, "Facing Up to the Problem of Consciousness!" Journal of Consciousness Studies, 2 (1995), 200-219.

[90] See footnote 16.

[91] There is a controversy about whether an AI has really passed the Turing Test. A majority of researchers seem to believe that it has, while a minority thinks that it is still out of reach for current AIs. For the argument made in this paper, however, this doesn't matter. Basically, any AI that is so "eloquent" that people get the impression that there might be a real "mind" behind it qualifies; for a discussion of the Turing Test, as well as competing concepts like the Lovelace Test or the Metzinger Test, see also: Knappe, 40.

[92] Steven Strogatz, "The Joy of Why: What Is the Nature of Consciousness?" Quanta Magazine, May 21, 2023, https://www.quantamagazine.org/what-is-the-nature-of-consciousness-20230531.

[93] For an illustration of the unbridgeable gap between human minds and those of differently structured organisms, see the classic 1974 paper by Nagel, who chose a bat or Martians to demonstrate this: Thomas Nagel, "What Is It Like to Be a Bat?" The Philosophical Review, Vol. 83 4/1974, 435-450.

# 6 AI Art Today and In the Future

## 6.1 Bold Claims Outstrip Tiny Technology: The Current State of Artificial Creativity in Artistic Projects[94]

In the last years, there have been many artistic projects in which the advent of AI as a creative "alien mind" has been claimed. This to the point that it appears to be a prevailing trend to elevate AI systems to the highest level of authorship in artistic endeavors.[95]

None of these projects qualify as true independent agent AIs. They often even lack a real collaborative element and the AIs employed primarily function as tools, similar to Photoshop filters, with their "independence" mainly based on guided randomness. The creativity in these projects is simplistic, combinational and in rare cases exploratory in nature, driven by basic interpolation. In the most technically advanced efforts, elements of "assisted creation" as an amplifier of human creativity can be seen at best. While some projects claim transformational creativity, like the CAN-project, the results remain strictly constrained within the boundaries of the training data due to the limitations in dataset size and the network's simplicity.[96]

---

94 This is only a brief and sketchy overview; for further reading, see the detailed and nuanced overview of the AI art scene, with numerous interviews with some of the most important artists until 2021: Anne Ploin et al.; for a brief and concise overview of the major developments in AI art up to 2021: Cetinic, 6-9; for those who can read German, issue no. 278 of the art magazine "Kunstforum" gives a good and insightful overview of the AI art scene and some of the issues that go along with it: Kunstforum International 278 (2021): Kann KI Kunst? AI Art: Neue Positionen und technische Ästhetiken; in particular, the articles by Pamela C. Scorcin, the late Peter Weibel, and Arthur I. Miller provide a good background on the crucial developments in AI art; for an overview of some of the artists' positions, see the anthology: Craig Vear et al., The Language of Creative AI: Practices, Aesthetics and Structures (Cham, 2022). For an overview of AI art and a discussion of how to use the technology as an artist, see: Sofian Audry. Art in the Age of Machine Learning /Cambridge (MA), 2021). For a discussion of AI as a collaborator, especially in the process of drawing, see: Frederic Fol Leymarie et al., "What Is It about Art? A Discussion on Art.Intelligence.Machine," Arts, 11/5 (2022), https://doi.org/10.3390/arts11050100.

95 The most prominent projects that present an AI as a fully autonomous agent are Stephen Thaler's "Creativity Machine". Thaler is involved in a number of lawsuits to claim copyright for the creations of his AI (see. Thom Whaite, "Can AI Ever Be Truly Creative? A New Court Case Lays Down the Law," Dazed and Confused, Aug 23, 2023, https://www.dazeddigital.com/art-photography/article/60654/1/can-ai-ever-be-creative-new-court-case-law-creativity-machine-stephen-thaler); Simon Colton's "Paintingfool" project, started in 2006, about an AI that "wants to be accepted as an artist in its own right" (see for a detailed description: Simon Colton, "The Painting Fool: Stories from Building an Automated Painter," in: Computers and Creativity , ed. Jon McCormack et al. (Cham, 2012), 3-38); also the "CAN" project by Ahmed Elgammal (see endnote 96) made the claim that the fictitious AI artist behind it, AICAN was an autonomous agent. This claim, and Elgammal's refusal to clarify that this AI is merely a tool, has been criticized, and a financial motive has been suggested, as AICAN art has been exhibited worldwide and individual works have sold for thousands of dollars (see: Ian Bogost, "The AI-Art Gold Rush is Here. An Artificial-Intelligence "Artist" Got a Solo Show at a Chelsea Gallery. Will It Reinvent Art, or Destroy It?" The Atlantic, March 6, 2019, https://www.theatlantic.com/technology/archive/2019/03/ai-created-art-invades-chelsea-gallery-scene/584134/). The very same motivation was assumed behind the claim that the simple GAN architecture that produced the "Edmond de Bellamy" artwork was also highly anthromorphized and marketed as an independent agent with the slogan "creativity is not just for humans" (see: Ziv Epstein et al., "Who Gets Credit for AI-Generated Art?" Iscience, 23/9/2020, doi: 10.1016/j.isci.2020.101515).

96 Ahmed Elgammal's "CAN" (Creative Adversarial Network) project was a system fed with 80,000 images of Western art from 1400 to 2000. The system was designed to "rate" the "degree of creativity" of each individual artwork, and independently create what was claimed to be a new, never-before-seen art style from all of them. The goal was to systematically produce novelty. See: Ahmed Elgammal et al., "CAN: Creative Adversarial Networks, Generating "Art" by Learning About Styles and Deviating from Style Norms," ArXiv, June 21, 2017, https://doi.org/10.48550/arXiv.1706.07068.

In these projects, the claim of an "alien mind" or "independent agent'" often carries a gimmicky and marketing tone.[97] It's a mix of overt anthropomorphization of even the simplest algorithms in combination with the elevation of simple consumer computer technology to an "alien mind". These bold claims simply outstrip the tiny technology that works under the hood.[98] The reason for these exaggerations is obvious: to create an independent agent AI would be nothing less than to produce the greatest work of art of all time: artificial life made from inanimate matter. And to make this bold claim is obviously almost irresistibly tempting.

All of this is intriguing, indicating the salience of this issue in the art world. There is a good deal of wishful thinking about true alien minds turning into genuine artists. While these projects and artworks are clearly no trade shows of AI technology, their "value" lies in their contextualization. They can be seen as "apparative conceptual art"[99] with the main value to initiate a discourse about these technologies.

Currently, the shift in authorship can be clearly detected, but this is in the use of these large systems, for example in image generators, when their output is considered in isolation and at face value. Modifying those systems can be done easily, and creative projects utilizing these new capabilities should be imminent. However, it requires the artist to step back from the entire process and to give the AI space for an independent act of creation. And in most projects the artists themselves still clearly stand in the foreground and do not seem ready for this altruistic act.

The sheer volume of AI art will increase. And as the body of this work grows and adds layer upon layer to the existing mass, its gravitational pull will increase to the point where it may begin to attract the type of attention, desire, and money that usually is only associated with the likes of Jeff Koons, Damien Hirst, Gerhard Richter, Ai Weiwei, or Beeple for that matter.[100] Early AI artworks, especially those using GANs, may be seen as prophetic in retrospect. And as the monetary worth attributed to these works increases, so does the general perception of value and importance in the art world.

**6.2 More Than Human Worlds: Challenging Human Narcissism and Anthropocentrism**

Looking at and into "alien minds" might fundamentally challenge humanity's self image. Some AI art is already tapping into "more than human worlds"[101] to undermine our - as these artists phrase it - often unreflected and naive anthropocentric perspective.[102] This aligns with a broader scientific movement attributing cognition and intelligence to various organisms, not just humans and other mammals.

---

97 See also on the marketing gag character of most of these works: Anna Notaro, "State of the Art: A.I. Through the (Artificial) Artist's Eye," Proceedings of EVA London 2020, https://www.scienceopen.com/hosted-document?doi=10.14236/ewic/EVA2020.58.

98 For a critique of such "misleading" claims, see: Aaron Hertzmann, "Computers Do Not Make Art, People Do," Communications of the ACM 63, 5/2020, 45–48, https://dl.acm.org/doi/10.1145/3347092.

99 Pamela C. Scorzin uses this term ("apparative Konzeptkunst") in reference to the 2016 project "The Next Rembrandt," in which an AI was trained on Rembrandt paintings to produce something like a "typical" work of art by the Flemish master. She detects an element of true "computational creativity" in the overall contextualization of this project between technology, marketing and the public, see: Pamela C. Scorzin, "ARTificiality: Künstliche Intelligenz, Kreativität und Kunst," Kunstforum International, 278/2021, 50-75, 69-70.

100 Beeple (Mike Winkelmann) is a self-taught CGI artist who has been posting one digital image a day on his social media channels for over 10 years, and in February 2021, he dropped 5000 images as an NFT that fetched nearly $70 million at a Christie's auction. This was the peak of the NFT art craze in the midst of the Corona cryptocurrency boom, see also: Jacob Kastrenakes, "Beeple Sold an NFT for $69 Million," The Verge, May 11, 2021, https://www.theverge.com/2021/3/11/22325054/beeple-christies-nft-sale-cost-everydays-69-million.

101 Carlos Castellanos et al., "Beauty: A Machine-Microbial Artwork," EasyChair Preprint, April 30, 2021, 4.

102 Or as the artist duo Cesar & Lois put it: "anthropocentric orientations of intelligence pervade contemporary art and popular culture," see: Cesar & Lois et al., "Ecosystemic Thinking: Beyond Human Narcissism in AI," in: The Language of Creative AI: Practices, Aesthetics and Structures, ed. Craig Vear et al. (Cham, 2022), 95-111, 95.

An artwork that falls into this category is one that exposes the microscopic worlds that are usually hidden from our view, systematically exploiting the generative capabilities of bacteria controlled by an AI that sees this microcosm as its world.[103] This project, aptly named "Beauty," showcases pattern-plotting stems of bacteria that produce just that, and all claims about the AI employed are technologically fully justified.

Humanity's notion of absolute exceptionalism may be increasingly being challenged by science. Aesthetically, AI art may systematically attack what has already been identified in this context as specism and anthropocentric "human narcissism."[104] By design, AI may be the ideal collaborative partner for such projects, and Artificial Creativity its engine.

**6.3 Melting With the Machine – Tapping Into Unlimited Creativity**

When it comes to AI scenarios of a more remote future, when things get more outlandish and science-fictionesque, one idea almost inevitably enters the picture: the vision of transhumanism. It emerged in the 1950s, parallel to the establishment of AI as an independent field of study,[105] and gained prominence in the 1990s with protagonists like Marvin Minsky, Ray Kurzweil, and Hans Moravec.[106] This movement stirred controversy due to its proximity to eugenics.

Arthur I. Miller has taken up this scenario, stripped it of its negative implications and sees in this the future of Artificial Creativity and AI art. He envisions AI producing "unlimited creativity" that humans can only become part of when they are ready to technologically enhance their minds.[107] The argument goes like this: As things progress, AI will become so overtly imaginative and inventive for itself that it will start to produce something like "out-of-date" artifacts and artworks, something like the ancient Greek *Antikythera* mechanism or works like Joyce's "Finnegan's Wake" or Lou Reed's "Metal Machine Music," and this in rapid succession. Ordinary people will not be able to understand or value this, and might even be repelled by it all, but the tiny avantgarde of enhanced humans will be able to appreciate this kind of Artificial Creativity. And over time, this – along with other pull factors – will attract more and more humans to become cyborgs. Eventually, merging with machines will seem inevitable. In this future, art created by biological humans could lose its relevance and become obsolete. Only the art of the enhanced will truly speak to the physically augmented. And this would also be the stage at which AI would finally have acquired true independent agency and what is left of biological humanity would be a collaborative sparring partner at best.

---

103 It literally makes use of an open sourced "world model" AI, see for a description of the art project: Carlos Castellanos, "Intersections of Living and Machine Agencies: Possibilities for Creative AI," in: The Language of Creative AI: Practices, Aesthetics and Structures, ed. Craig Vear et al. (Cham, 2022), 155-166.
104 Cesar & Lois, 95.
105 Julian Huxley, "Transhumanism", New Bottles for New Wine (London, 1957), 13–17.
106 Minsky was the co-founder of the AI lab at MIT and one of the world's most prominent computer scientists. He was an initiator of the 1956 Dartmouth conference that launched AI as an independent field of research. Kurzweil was Minsky's student at MIT and became a best-selling popular science author. He can be considered a classic futurist and has been employed by Google. Moravec was a professor at Carnegie Mellon University and was the most vocal and controversial proponent of transhumanism in academia.
107 Miller, 311-313.

# 7 Conclusion and Outlook

The genie is out of the bottle. Since 2022, AI systems have gained generative capabilities that constitute a giant qualitative leap over all their predecessors. Now there is almost everything in the output of machines usually attributed to creativity in humans. With the help of generative AI, a mediocre, minimal, or almost non-existent creative human input can trigger a process that results in what appears to be an artistic masterpiece. In human-computer collaboration, the line of authorship has shifted toward the machine.

Technically, current pure ML architectures still limit the possibility of creating truly new, unprecedented instances which might entail the potential for revolution. But this hindrance can be eliminated by hybrid systems employing algorithms that systematically identify target areas for outputs outside the boundaries of their conceptual spaces. And thereby this, the highest form of what humans have ever been able to achieve in terms of creativity, could find standardization in machines.

While technology is evolving at a pace that even true machine agency shimmers at the horizon, in the art world this claim has been made since AI became available as part of the artist's toolkit. Today there is still a deep gap between this narrative and the true nature of these projects which mostly employ AI technologies as simple tools. It may take a while to adjust, but the potential for exciting art in this field is abundant.

In all of this, lies more than just new technology used in artworks. In fact, this is orders of magnitude bigger. Current AI art, with its bold claims made out of tiny technology, may in retrospect be seen as an interesting illustration of a transitional period of various foreshadowings, some of which may later crystallize into the form of factual reality. And today's interaction with generative AIs may very well be perceived as humanity's first serious contact with (the concept of) "alien minds." Still homemade, but alien.

However, as exciting this perspective may be, there is no way to find out if there is sentience and a "real" inner conscious experience behind this. Therefore, it may be wise to separate the question of machine authorship from that of consciousness, even though they are seemingly inextricably intertwined. And this issue also goes to the core of the perception and valuation of Artificial Creativity as such:

Can machines be truly creative? The way we interpret creativity always implies individual agency, which is associated with sentience or consciousness. The creative process is the outcome of internal reflections on human experience, feelings, and thoughts. This is what makes art meaningful to us. It is surrounded by myths which are deeply engrained in human thinking. Creativity is considered one of the highest qualities of a person, as the marvel of the human mind. And it is precisely this that makes the claim that there is such a thing as Artificial Creativity so controversial and unsettling to many, even more so than the idea of Artificial Intelligence as such. There lies an Uncanny Valley element in all of this. It is as if we are faced with a Frankenstein monster that we now begin to realize we have brought into existence.

However, the categorical rejection of the possibility of machine creativity entails an element of mystification of human genius. And it plays into what might be identified in the future as anthropocentric chauvinism. We have entered unsafe, slippery terrain in what was once rock-solid ground.